\documentstyle[epsf,a4]{article}  
\begin{document}                % INITIALIZE - DONT CHANGE
\newcommand{\manual}{rm}        % Substitute rm (Roman) font.
\newcommand\bs{\char '134 }     % add backslash char to \tt font
\newcommand{\Het}{$^3{\mathrm{He}}$}
\newcommand{\Hef}{$^4{\mathrm{He}}$}
\newcommand{\A}{{\mathrm{A}}}
\newcommand{\D}{{\mathrm{D}}}
\newcommand{\simlt}{\stackrel{<}{{}_\sim}}
\newcommand{\simgt}{\stackrel{>}{{}_\sim}}
\newcommand{\MeV}{\;\mathrm{MeV}}
\newcommand{\TeV}{\;\mathrm{TeV}}
\newcommand{\GeV}{\;\mathrm{GeV}}
\newcommand{\eV}{\;\mathrm{eV}}
\newcommand{\cm}{\;\mathrm{cm}}
\newcommand{\s}{\;\mathrm{s}}
\newcommand{\sr}{\;\mathrm{sr}}
\newcommand{\lab}{\mathrm{lab}}
\newcommand{\ts}{\textstyle}
\newcommand{\ol}{\overline}
\newcommand{\be}{\begin{equation}}
\newcommand{\ee}{\end{equation}}
\newcommand{\ba}{\begin{eqnarray}}
\newcommand{\ea}{\end{eqnarray}}
\newcommand{\rau}{\rho_{\mathrm Au}}
\newcommand{\nn}{\nonumber}
\newcommand{\N}{{\mathrm{N}}}
\newcommand{\css}{({\mathrm{cm}}^2-{\mathrm{s}}-{\mathrm{sr}})^{-1}}
\newcommand{\pp}{$\overline{p}(p)-p\;\;$}
\renewcommand{\floatpagefraction}{1.}
\renewcommand{\topfraction}{1.}
\renewcommand{\bottomfraction}{1.}
\renewcommand{\textfraction}{0.}               
\renewcommand{\thefootnote}{F\arabic{footnote}}
\title{Stronger Neutrino Interactions at Extremely
High Energies and the Muon Anomalous Magnetic Moment}
\author{Saul Barshay and Georg Kreyerhoff\\
III. Physikalisches Institut\\
RWTH Aachen\\D-52056 Aachen\\Germany}
\maketitle
\begin{abstract}                % DON'T CHANGE THIS LINE
A specific model of parity-conserving lepton substructure
is considered. We show that a positive-definite contribution
to the muon $(g-2)/2$ at the possible level of about
$4\times 10^{-9}$, can be related to a significant increase
in the interaction cross section for cosmic-ray neutrinos
with energies above about $10^{19}$ eV. The additional cross
section at $\sim 10^{20}$ eV is calculated to be $\sim 10^{-29}$ cm$^2$,
which is about 100 times the standard weak-interaction cross
section. The model involves an extremely massive, neutral
lepton, with $m_L\cong 2\times 10^6$ GeV fixed by the new
contribution to $(g-2)/2$.
\end{abstract}
One may consider a recently measured small deviation\cite{ref1}
from the theoretically-calculated value of the muon  anomalous
magnetic moment as an indication of possible new physics.\cite{ref2} In doing so, it would be valuable if two conditions were fulfilled.
\begin{itemize}
\item[(1)] The sign of the experimental deviation is positive. The
new physics should produce this sign unambiguously. That is, it
should contribute to the anomalous magnetic moment a positive-definite
quantity.
\item[(2)] It would be useful if the new physics implied other
definite physical processes which are directly calculable. Further,
these processes should be accessible to exploration in current
experiments.  They may well be related to puzzling aspects of the
present experimental situation in different areas. 
\end{itemize}
In this paper, we give and discuss the results of straight-forward
calculations based upon hypothetical additional\footnote{
That is, additional to the dynamics contained in QED and in the
standard electroweak-QCD model.
}
dynamics which relates the possible deviation in $(g-2)/2$ for
the muon to the possible occurrence of stronger-than-weak interactions
for muon (and tau) neutrinos at very high energies, $E_\nu > 10^{19}$ eV.
The latter possibility concerns the nature and origin of the particles
which initiate the highest-energy cosmic-ray air showers, with energies
up to about $10^{20}$ eV. This issue will be under further intense
study in on-going\cite{ref3,ref4} and up-coming cosmic-ray experiments
\cite{ref5,ref6}.\par
In the early classic review of QED\cite{ref7}, a variety of contributions
to the muon $(g-2)/2$ were calculated from hypothetical couplings to
``exotic particles'' (section  7.3). A positive-definite result\cite{ref7}
is obtained for the process shown in the Feynman diagram in Fig.~1,
\be
\Delta a_\mu = \frac{g^2}{4\pi}\frac{1}{4\pi}\left( \frac{m_\mu}{m_L}\right)
\left\{ 1 - \frac{2}{3}\left(\frac{m_\mu}{m_L}\right)\right\},\;\;\;\;
m_L\gg m_\mu \sim m_\pi
\ee
In Fig.~1, $L$ denotes a very massive, spin-1/2 (muonic) neutral lepton
which is assumed to be coupled to a muon and a hypothetical \underline{pointlike component}
of a pion ( a Goldstone component? )  with a
\underline{strong parity-conserving}, effective Hermitian interaction
of the form
\be
ig\left\{(\ol{L}\gamma_5\mu)\pi^+ + (\ol{\mu}\gamma_5 L)\pi^-\right\},
\;\;\; \gamma_5^\dagger=\gamma_5,\; g\;\mbox{real}
\ee
where the particle symbols stand for the fields. With $g^2/4\pi\sim 1$
and\cite{ref1} $\Delta a_\mu \sim 4\times 10^{-9}$, Eq.~(1) immediately
gives a very large mass,
\be
m_L\cong 2\times 10^6 \GeV
\ee
The general physical idea\cite{ref8} embodied phenomenologically in Eq.~(2)
is that of a parity-conserving substructure in leptons, a structure
which is very compact in space because of an effectively very massive
``core'' region. It is not necessary to augment this idea with specific
theoretical details in order to go immediately to a further unusual
new consequence for certain experiments at the highest energies.
Consider the analogue of Eq.~(2) involving a neutrino $\nu_\mu$,
\be
ig\left\{ (\ol{L}\gamma_5\nu_\mu) \pi^0 + (\ol{\nu}_\mu\gamma_5 L)\pi^0\right\}
\ee
For a center-of-mass energy $\sqrt{s}\cong m_L\cong 2\times 10^6$ GeV
in a neutrino-nucleon collision in the atmosphere, the neutrino energy
must be $E_\nu \sim \left(s/2m_\N\right)\cong 2\times 10^{21}$ eV.
This energy is a factor of about 10 above that of the present small
number\cite{ref3,ref4} of highest-energy cosmic-ray air showers,
which are estimated to be at, or just above, $10^{20}$ eV. However,
the process\footnote{
An analogous process involves an extremely high-energy incident muon.
}
shown in the Feynman diagram in Fig.~2 gives rise to
an additional neutrino interaction cross section in the atmosphere.
Here the $L$ contributes virtually; the final-state products are
a lepton and a pion. In addition, there is the multi-hadron production
from the pion total absorption cross section which controls the strength
of the squared lower vertex from Fig.~1. What is particularly interesting
is the result of our calculation of the additional neutrino cross
section, shown in the graph in Fig.~3. For $m_L\cong 2\times 10^6$ GeV
(fixed by $\Delta a_\mu$ with a strong $g^2/4\pi\sim 1$),
$\Delta \sigma_\nu(\sqrt{s})$ is comparable to a standard weak-interaction
cross section of the order of $10^{-31}$ cm$^2$ at $E_\nu\sim 10^{19}$ eV,
but steadily rises to a considerably larger value of $\sim 2.7\times 10^{-29}$
cm$^2$ at $E_\nu\sim 2\times 10^{20}$ eV.  This rise is of course,
just what is expected for the approach to production of real $L$, as
$E_\nu$ increases (but possibly never reaches  $\sim 2\times 10^{21}$ eV,
for actual cosmic-ray neutrinos\cite{ref9}). The expression for
$\Delta\sigma_\nu(\sqrt{s})$ giving the curve in Fig.~3 is (in millibarns),
\ba
\Delta \sigma_\nu(\sqrt{s})&=& \left(\frac{g^2}{4\pi}\right)^3
\frac{4\sigma^{{\mathrm{tot}}}_{\pi\N}}{\pi}\int_0^{\frac{\sqrt{s}}{2}}
dL (L^2)\int_0^\frac{(s-2\sqrt{s}L)^{1/2}}{2} dQ (Q^2)\gamma\nn\\
&\times&
\int_{-1}^1 dx \frac{\left(\frac{s}{4}+L^2-\sqrt{s}Lx\right)^{1/2}
(\gamma 2 Q - Lx)}{(4Q^2-m_L^2)^2(4Q^2-\sqrt{s}(\gamma2Q-Lx)-m_\pi^2)^2}\nn\\
&\times& \frac{(1-v)}{v}\ln \left(\frac{1+v}{1-v}\right)
\ea
with $\gamma=1/(1-v^2)^{1/2}$, $\;\;v=L/(4Q^2+L^2)^{1/2}$,
$\;\;\;m_L=2\times 10^6$ GeV.\par\noindent
The cross section $\sigma_{\pi\N}^{\mathrm{tot}}$ is approximated by a
constant of the order of 200 mb.\cite{ref10}. The squared invariant mass in the final
lepton-pion system is $4Q^2$, $L$ is the magnitude of the three-momentum
of the virtual, massive lepton; this three-momentum makes an angle
$\theta$ ($x=\cos\theta$) with the incident neutrino direction in the
c.~m.~system of the neutrino-nucleon collision. Masses of nucleon and
final lepton and pion are neglected in the kinematics; the pion mass
is included in the pion propagator to formally avoid singular behavior.
Clearly, with an interaction cross section of the order of 100 times
greater than that from the electroweak interaction, cosmic-ray neutrinos
with $E_\nu\sim 10^{20}$ eV have a significantly greater probability
to produce detectable air showers, if there is sufficient flux.
However, $\Delta\sigma_\nu$ is probably still not large enough to
account for the (presently  few, $\sim 14$) highest-energy events\cite{ref4} 
coming from small zenith angles\cite{ref3} ($<45^\circ$). This might 
be possible if there is a very high energy neutrino flux as large \footnote{
A flux of $\tau$ neutrinos of up to about $10^{-16}\css$ has been
estimated in \cite{ref9,ref12} as arising from the 
\underline{two-body} decay $\phi\to\nu_\tau\ol{\nu}_\tau$, of a 
metastable, scalar inflaton (lifetime of $\sim 10^{26}$s because
the decay amplitude is proportional to the neutrino mass). The
$\phi$ mass has been calculated, from the minimum of the potential,
to be of the order of $10^{11}$ GeV. \cite{ref12,ref13}. The $\phi$
can constitute a significant part of dark matter throughout
the present universe.\cite{ref12}
}
\cite{ref9} as about $10^{-17}\css$, and an interaction probability
in air as large as about $10^{-3}$. But there must be air showers at
$10^{20}$ eV coming from large zenith angles. Recently, two such showers have
been reported and analyzed.\cite{ref11} The study of the frequency
and of the characteristics of air showers at large zenith angles
is expected to increase. The Auger air-shower array\cite{ref5} will
augment the present sensitivity for ``horizontal'' neutrino-induced air
showers at $E_\nu\sim 10^{20}$ eV, down to a possible flux of 
about $10^{-17}\;\css$. The Antarctic ice detector specifically for neutrinos
(and muons), 
when in expanded operation, is said\cite{ref6} also to be able to detect a
flux of neutrinos as low as $10^{-17}\css$ at $E_\nu\sim 10^{20}$ eV.\par
There is a further unusual physical process that can be envisioned,
and estimated in the context of the above ideas\footnote{
Note that a contribution to the electron $(g-2)/2$ of about
$(m_e/m_\mu)\frac{1}{2}(\Delta a_\mu)\sim (1/400)(\Delta a_\mu)$ is
marginally allowed by the present uncertainty (i.~e.~via an electron
$L_3$ with $m_{L_3}\sim 2m_L$). But absence of $\mu\to e\gamma$,
requires essentially no mixing of $L_3$. 
}
: this is
$\tau\to \mu+\gamma$. Consider two $L$ particles, $L_1$ and $L_2$ with 
masses $m_{L_1}$ and $m_{L_2}$, which mix with an angle $\theta_L$ in
the flavor states
\ba
L_\mu = (L_1\cos\theta_L + L_2\sin\theta_L)\nn\\
L_\tau =  (L_2\cos\theta_L - L_1\sin\theta_L)
\ea
The effective interaction analogous to Eq.~(4) is
\be
ig\left\{(\ol{L}_\mu \gamma_5\nu_\mu)\pi^0 + 
(\ol{L}_\tau \gamma_5\nu_\tau)\pi^0 + h.c.\right\}
\ee
The branching ratio for $\tau\to\mu+\gamma$ is then,
\be
B.R.(\tau\to\mu+\gamma)\cong\frac{\frac{\alpha}{4}|\tilde{\Delta a_\mu}|^2m_\tau}
{(\tau_\tau)^{-1}}\sim 1.5\times 10^{-9}
\ee
where the $\tau$ lifetime is $\tau_\tau\cong 2.9\times 10^{-13}$ s, the
mass is $m_\tau\cong 1.78$ GeV, and $\alpha=1/137$. The quantity 
$|\tilde{\Delta a_\mu}|$ is
\be
|\tilde{\Delta a_\mu}|\sim (m_L\Delta a_\mu)\times \left|
(\cos\theta_L\sin\theta_L)\frac{(m_{L_2}-m_{L_1})}{m_{L_1}m_{L_2}}\right|
\sim \frac{\Delta a_\mu}{4}
\ee
where we have \underline{assumed}, simply for illustration,
$|(\cos\theta_L\sin\theta_L)m_L(m_{L_2}-m_{L_1})/(m_{L_1}m_{L_2})|$ to
be of the order of $1/(2)^2$ i.~e.~for non-zero $|\theta_L|\sim \pi/4$,
and with $m_{L_2}\sim \frac{m_{L_1}}{1.5} = \frac{m_L}{1.5}$.$^{F4}$ 
The present experimental upper limit for this branching ratio is $1.1\times
10^{-6}$.\par
In summary, a strong, parity-conserving interaction of a colorless
quark composite like a pion with a muon and a very massive
neutral lepton, gives a positive-definite contribution to the
muon anomalous magnetic moment of about $4\times 10^9$, when the
massive lepton mass $m_L$ is about $2\times 10^6$ GeV. An analogous
interaction for neutrinos\footnote{
Other possible effects, such as an induced, neutrino (and $L$) magnetic
moment ($O(10^{-11}\mu_B)$), although interesting, appear to be well below
present experimental upper limits.
} leads directly to an additional
interaction cross section for very high energy cosmic-ray neutrinos in
the atmosphere. As a result of the above explicit value of $m_L$ extracted
from the present discrepancy in $a_\mu$\cite{ref1}, and the
explicit dynamics in Fig.~2, computation gives a $\Delta\sigma_\nu(\sqrt{s})$
which surprisingly, becomes comparable to the standard weak-interaction
cross section of about $10^{-31}$ cm$^2$ at $E_\nu=10^{19}$ eV. The
calculated result (Fig.~3) that $\Delta\sigma_\nu(\sqrt{s})$ increases
strongly\footnote{
The sensitivity to $m_L$ in Eq.~(5) is such that $m_L\to m_L/1.5$,
increases the $\Delta\sigma_\nu$ in Fig.~3 by about 5. With exclusive
reference to $\tau$ neutrinos, an $m_{L_2}\sim m_L/3$ yields a 
$\Delta\sigma_{\nu_\tau}\sim 10^{-27}\cm^2$ at $\sim 10^{20}$ eV;
the atmospheric interaction of $\nu_\tau$ would become hadron-like.},
to about 100 times $\sigma_{\mathrm{weak}}$ at $E_\nu\sim 10^{20}$
eV, opens the possibility for an observable neutrino-induced component
in the highest-energy cosmic-ray air showers, if the neutrino flux
is as high\cite{ref9} as $10^{-17}\css$. Thus, experiments\cite{ref1,ref4,ref5,ref6}
on the muon
$(g-2)/2$ and on the interaction of extremely high-energy neutrinos, may
be considered as related probes for a definite kind of lepton substructure at 
extraordinarily small distances.\footnote{
In this model with $\gamma_5$ coupling, a radiative correction can give
a \underline{downward} shift from a ``bare'' lepton mass.\cite{ref2}
}

\newpage
\section*{Figures}
\begin{figure}[h]
\begin{center}
\mbox{\epsfysize 8cm\epsffile{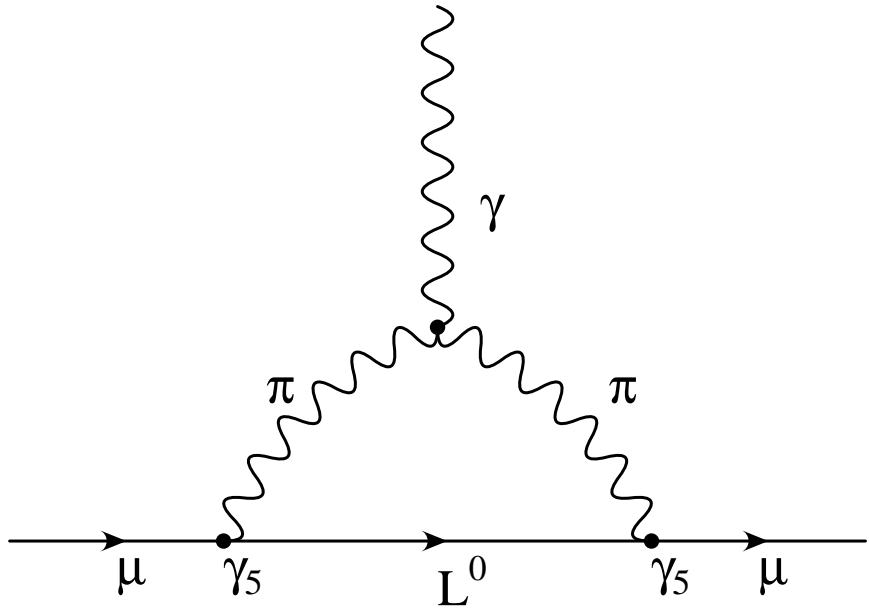}}
\caption{Feynman diagram for a contribution to the muon anomalous magnetic
moment due to a strong, parity-conserving interaction with a very
massive, neutral lepton $L^0$.}
\end{center}
\end{figure}
\begin{figure}[b]
\begin{center}
\mbox{\epsfysize 8cm\epsffile{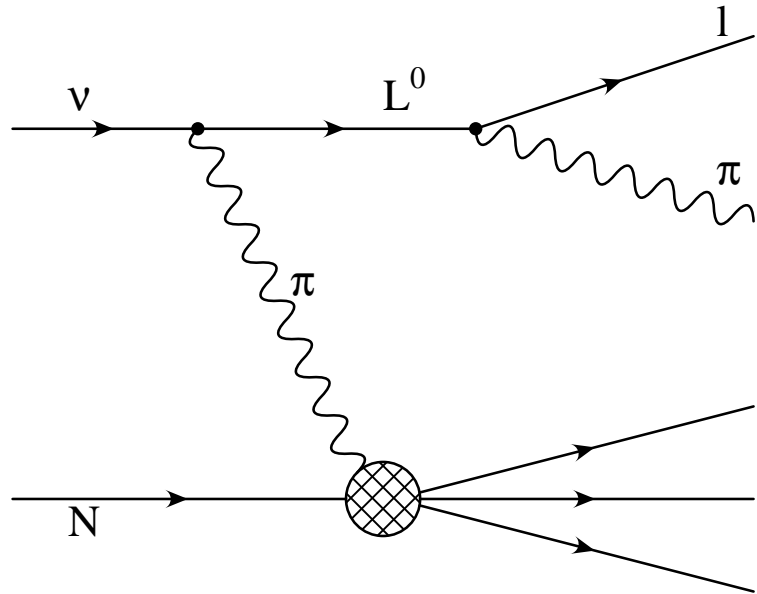}}
\caption{Feynman diagram for a contribution to the interaction cross
section of an extremely high energy, cosmic-ray neutrino with an 
atmospheric nucleon, mediated by a virtual $L^0$.}
\end{center}
\end{figure}
\begin{figure}[t]
\begin{center}
\mbox{\epsfysize 12cm\epsffile{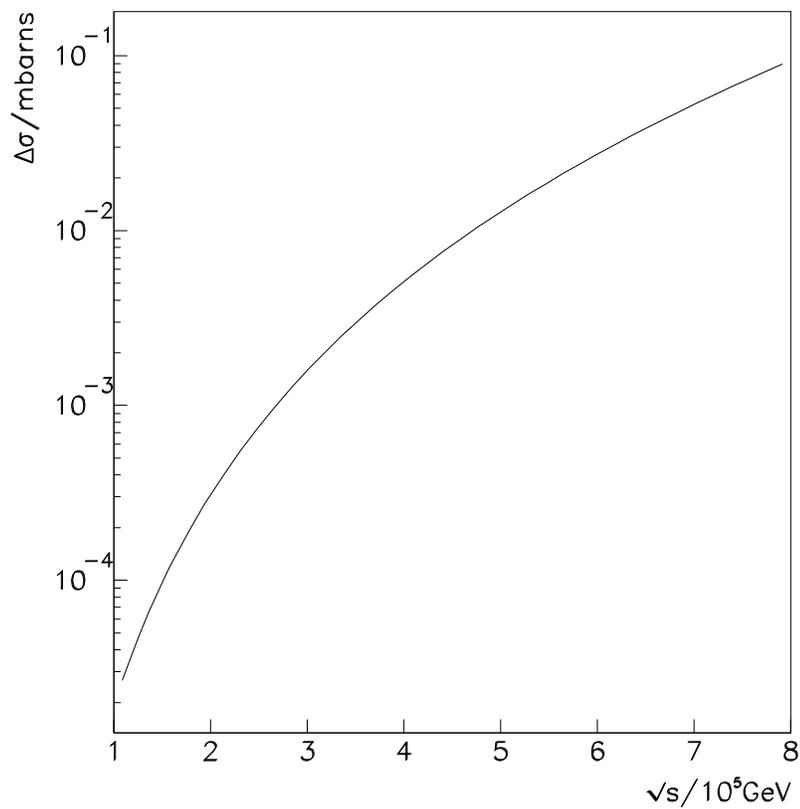}}
\caption{The additional neutrino-nucleon interaction cross section 
$\Delta\sigma(\sqrt{s})$ calculated from Eq.~(5), plotted versus the 
c.~m.~energy $\sqrt{s}$. The curve is for cosmic-ray neutrinos with
$E_\nu$ from $\sim 10^{19}$ eV to $\sim 3\times 10^{20}$ eV.}
\end{center}
\end{figure}
\end{document}